\documentclass{pasj00}
\usepackage{graphicx,lfootnote,enumerate, lscape}
\renewcommand{\thefootnote}{\alph{footnote}}

\begin{document}
\SetRunningHead{Iye et al.}{Subaru  Meteor Observation}
\Received{2006/12/20}

\title{SuprimeCam Observation of Sporadic Meteors during Perseids 2004}

 \author{
Masanori \textsc{Iye}\altaffilmark{1-3},
Mikito \textsc{Tanaka}\altaffilmark{2},
Masahisa \textsc{Yanagisawa}\altaffilmark{4}, \\
Noboru \textsc{Ebizuka}\altaffilmark{5},
Kouji \textsc{Ohnishi}\altaffilmark{6},
Chikako \textsc{Hirose}\altaffilmark{7} \\
Naoko \textsc{Asami}\altaffilmark{2},
Yutaka \textsc{Komiyama}\altaffilmark{1},
and
Hisanori \textsc{Furusawa}\altaffilmark{8}
}
\altaffiltext{1}{Optical and Infrared Astronomy Division, National Astronomical 
Observatory, 
Mitaka, Tokyo, 181-8588 Japan}
\email{iye@optik.mtk.nao.ac.jp}
\altaffiltext{2}{Department of Astronomy, School of Science, University of Tokyo, 
Tokyo, 113-0033 Japan}
\altaffiltext{3}{Department of Astronomy, Graduate University for Advanced Studies, 
Hayama, Kanagawa 240-0193 Japan}
\altaffiltext{4}{The University of Electro-Communications, Chofugaoka, 1-5-1, Chofu, 
182-8585 Tokyo Japan}
\altaffiltext{5}{RIKEN, Hirosawa 2-1, Wako, 340-0042 Saitama, Japan}
\altaffiltext{6}{Nagano National College of Technology, Tokuma 716, Nagano, 381-8550, 
Japan}
\altaffiltext{7}{Consolidated Space Tracking and Data Acquisition Department, \\
Japan Aerospace Exploration Agency (JAXA), 2-1-1 Sengen, Tsukuba, Ibaraki, 305-8505, 
Japan}
\altaffiltext{8}{Subaru Telescope, National Astronomical Observatory of Japan, 
650 North A'Ohoku Place, Hilo, HI 96720 USA}

 
\maketitle

\begin{abstract}
We report the serendipitous findings of 13 faint meteors and 44 artificial space objects by Subaru SuprimeCam imaging observations during 11--16 August 2004.  The meteors, at about 100km altitude, and artificial satellites/debris in orbit, at 500km altitude or higher, were clearly discriminated by their apparent defocused image sizes. 

 CCD photometry of the 13 meteors, including 1 Perseid, 1 Aquarid, and 11 sporadic meteors, was performed. We defined a peak video-rate magnitude by comparing the integrated photon counts from the brightest portion of the track traversed within 33 milliseconds (ms) to those from a 0-mag star during the same time duration.  This definition gives magnitudes in the range $4.0 < V_{vr} < 6.4$ and $4.1 < I_{vr}<5.9$ for these 13 meteors.  The corresponding magnitude for virtual naked-eye observers could be somewhat fainter especially for the V-band observation, in which the [OI] 5577 line lasting about 1 sec as an afterglow could contribute to the integrated flux of the present 5--10 min CCD exposures.

 Although the spatial resolution is insufficient to resolve the source size of anything smaller than about 1 m, we developed a new estimate of the collisionally excited column diameter of these meteors.   A diameter as small as a few mm was derived from their collisionally excited photon rates, meteor speed, and the volume density of the oxygen atoms at the 100km altitude.  The actual column diameter of the radiating zone, however, could be as large as few 100m because the excited atoms travel that distance before they emit forbidden lines in 0.7 sec of its average lifetime.

 Among the 44 artificial space objects, we confirmed that 17 were cataloged satellites/space debris.  This shows the usefulness of SuprimeCam wide-field imaging programmed to study the faint meteor population and faint space debris.
\end{abstract}

\section{Introduction}
Scientific observation of meteors is not a simple task because of the difficulty in predicting their appearance in time and position.  The brightness of a meteor depends on the size and the velocity of the meteoroid.  Typical meteors seen by the naked eye or by video cameras are caused by meteoroids larger than about 0.01mm \citep{cepl98}.  However, the luminosity distribution of meteors at its fainter end is not well-known.  Moreover, the scientific definition of the magnitude of meteors is not well established because of the variety of parameters involved in making observations, for example, the speed and distance of the meteor and the spatial and spectral resolution and the spatial and spectral coverage of the instrument used, which can be the naked-eye, a video camera, an intensified camera, a CCD camera, or another device.

\citet{hawk58} and \citet{cook62} evaluated the width of meteor trails by comparing the measured trail width on photographs with those of stars and derived a typical value in the order of 1m.  Note that this evaluation is from broadband images and the decomposition analysis is a rather delicate process, as indicated by some nominal negative values that were derived in \citet{hawk58}. \citet{kai04} made two station, short-baseline ($\le$ 100m) high resolution measurements of 34 faint meterors to evaluate the trail width using intensified CCD cameras and found that their FWHM were generally less than 1m.  \citet{jenn04} serendipitously obtained an extended spectrum of a meteor trail with FORS1 on VLT but the apparent extension turned out to be caused by the defocus effect and the meteor was not spatially resolved.

The present paper reports the serendipitous observation of meteors made with the wide-field SuprimeCam of the 8.2m Subaru Telescope during the Perseid meteor shower of 2004.  

 Perseid meteors are known to be associated with the parent comet 109P/Swift--Tuttle , which was first recorded in BC 68 and returns every 135 years with recent visits in 1862 and 1992. The perihelion of 109P/Swift--Tuttle remained outside the Earth's orbit for at least for last 2000 years, but the current perihelion is at 0.958 AU. Its orbit is highly eccentric $e=0.963$ and highly inclined $i=113^{\circ}.4$ with the velocity relative to the Earth as large as about 60 km/s \citep{lyyt04a}. When it comes close to the Sun, a new trail is produced; at least five such dust trails are known from passages corresponding to AD 1348, 1479, 1610, 1737, and 1862.  
Taking the perturbation of Jupiter, Saturn and other planets into account, \citet{lyyt04} calculated that the one-revolution dust trail from the 1862 encounter passed within 0.0013 AU from the Earth's orbit at about 21 UT on 11 August 2004.

\section{Observations and Data Reduction}
The data used were obtained serendipitously during observations on other projects. Broadband imaging observations of M31 and narrowbands imaging observations of the Subaru Deep Field were made with the SuprimeCam \citep{miya02} attached at the prime focus of the Subaru Telescope \citep{iye04} during four nights, 12--15 August 2004, Hawaiian Standard Time (HST=UT-10), shortly after the expected epoch of Perseid meteor events. 

During the course of these observations, we serendipitously noticed several tracks recorded on the CCD images. Although no elaborate statistical assessment of the mean rate of meteor/artificial satellite tracks recorded on SuprimeCam was available at that time and although we didn't know how to securely discriminate meteors and artificial satellites without having double--station observations, we suspected that the event rate observed during this period could be significantly larger than the average.  Although we recognized that some of these recorded tracks must have been artificial satellites, we took a closer look. We asked SuprimeCam observers on the preceding four nights, 8--11 August, including the Perseid peak epoch, to check their images as well; one of these observers, I. Iwata, reported finding nine tracks in his images. Two other observers who conducted SuprimeCam imaging did not find such tracks during this period.  

\begin{figure}[t]
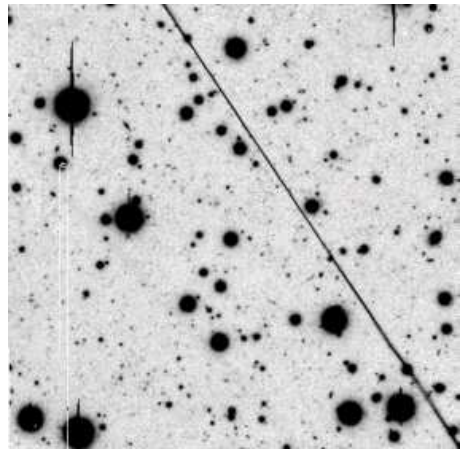

  \begin{center}
    \FigureFile(60mm,60mm){fig1.eps}
  \end{center}
  \caption{Track $\#11\_812\_39\_8$ \label{track11}}
\end{figure}

\begin{figure}
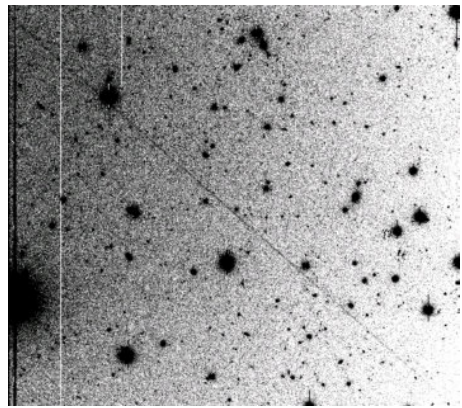

  \begin{center}
    \FigureFile(60mm,60mm){fig2.eps}
  \end{center}
  \caption{Track $\#50\_813\_56\_5$ \label{track50}}
\end{figure}

\begin{figure}
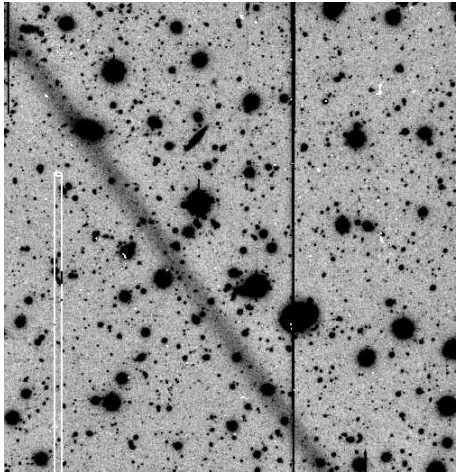

  \begin{center}
    \FigureFile(60mm,60mm){fig3.eps}
  \end{center}
  \caption{Track $\#17\_812\_69\_8$ \label{track17}}
\end{figure}

\begin{figure}
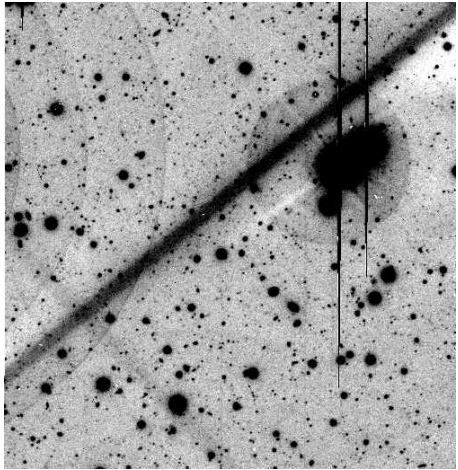

  \begin{center}
    \FigureFile(60mm,60mm){fig4.eps}
  \end{center}
  \caption{Track $\#30\_814\_46\_9$ (brighter one) and $\#57\_814\_46\_9$ (fainter 
one)\label{track30}}
\end{figure}

\begin{figure}
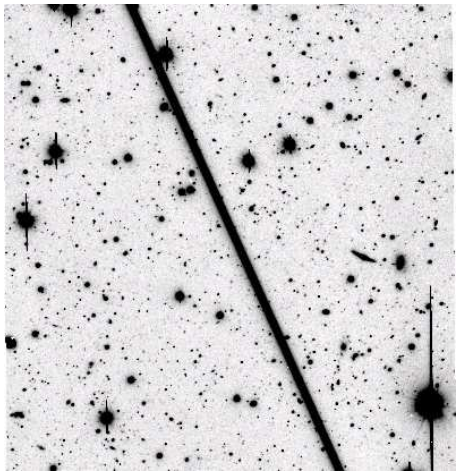

  \begin{center}
    \FigureFile(60mm,60mm){fig5.eps}
  \end{center}
  \caption{Track $\#39\_815\_55\_5$ \label{track39}}
\end{figure}

\begin{table*}

\small
  \begin{tabular}{rrrrrrrrrrrrrrr}
\\
\multicolumn{1}{c}{(1)} & \multicolumn{1}{c}{(2)} & \multicolumn{1}{c}{(3)} & 
\multicolumn{1}{c}{(4)} & \multicolumn{1}{c}{(5)}& \multicolumn{1}{c}{(6)} & 
\multicolumn{1}{c}{(7)} & \multicolumn{1}{c}{(8)} & \multicolumn{1}{c}{(9)} & 
\multicolumn{1}{c}{(10)} & \multicolumn{1}{c}{(11)} & \multicolumn{1}{c}{(12)} & 
\multicolumn{1}{c}{(13)} & \multicolumn{1}{c}{(14)} & \multicolumn{1}{c}{(15)}
\\
\hline
\# &ExpStart&Fr&CCDFr&RA&Dec&Az&El&Fil&Exp&$l$&$\theta$&$w$&$I_p$&$m_{nom}$\\
&HST h m s&No.&SUPA00-&deg&deg&deg&deg&&sec&pix&deg&$"$&ADU&\\
\hline
1 &08/11 04 03 58&-&-&14.0&12.5&-&-&V&600&2873&11.8 &-&-&\\
2 &08/11 04 14 58&-&-&14.0&12.5&-&80.0&V&600&4235&-16.4&-&-&\\
3 &08/12 01 53 08&-&-&14.0&12.5&-&-&V&600&2683&48.0&-&-&\\
4 &08/12 02 04 08&-&-&14.0&12.5&-&-&V&600&2004&77.8&-&-&\\
5 &08/12 02 26 10&-&-&14.0&12.5&-&-&V&600&1993&71.6&-&-&\\
6 &08/12 03 10 13&-&-&14.0&12.6&-&-&V&600&2186&-65.7 &-&-&\\
7 &08/12 03 32 14&-&-&14.0&12.6&-&-&V&600&3994&-0.7 &-&-&\\
8 &08/12 03 32 14&-&-&14.0&12.6&&81.0&V&600&4082&-12.8&-&-&\\
9 &08/12 03 32 14&-&-&14.0&12.6&&81.0&V&600&2353&-56.6 &-&-&\\
10 &08/13 00 23 14&38&337990&15.0&38.9&55.1&41.2&V&600&2801&-13.8 &1.0&120&19.6\\
11 &08/13 00 34 14&39&338000&15.0&38.9&54.7&43.3&V&600&7430&33.3&2.6&380&17.4\\
12 &08/13 01 08 12&42&338030&19.3&38.7&53.0&49.5&V&600&6462&12.1&1.4&500 &17.6\\
13 &08/13 03 42 05&64&338250&15.0&38.9&6.5&70.8&Ic&300&2178&23.6&1.0&120&18.4\\
14 &08/13 03 48 27&65&338260&14.0&39.2&0.3&70.6&Ic&300&2773&-39.5&1.2&60&19.1\\
15 &08/13 03 54 27&66&338270&14.0&39.2&356.8&70.5&Ic&300&3363&23.5&1.2&3200&14.7\\
16 &08/13 04 06 29&68&338290&14.0&39.2&349.9&70.2&Ic&300&5218&-31.9&1.6&1800&15.1\\
17 &08/13 04 12 30&69&338300&14.3&39.2&346.6&69.9&Ic&300&2793&34.8&13.6&38&16.9\\
18 &08/13 04 54 40&76&338370&7.6&38.1&318.2&63.4&Ic&300&6368&-33.3&2.6&1250&15.0\\
19 &08/14 00 42 21&37&338900&13.6&39.5&53.0&46.5&V&600&6609&22.9&12.2&195&16.4\\
20 &08/14 00 53 22&38&338910&13.6&39.5&52.0&48.6&V&600&6386&45.8&1.4&420&17.9\\
21 &08/14 01 37 29&42&338950&12.9&39.9&2.0&56.8&V&600&4057&-61.3&1.4&2000&16.2\\
22 &08/14 01 37 29&42&338950&12.9&39.9&46.0&56.8&V&600&7191&29.0&1.4&480&17.7\\
23 &08/14 03 26 47&59&339120&11.9&39.1&5.7&70.6&Ic&300&4189&47.9&13.4&55&16.5\\
24 &08/14 03 44 55&62&339150&7.3&37.8&343.6&71.1&Ic&300&5060&-72.6&1.6&800&15.9\\
25 &08/14 04 33 24&70&339230&17.3&37.6&338.4&70.6&Ic&300&7533&-65.5&13.4&120&15.7\\
26 &08/14 20 18 11& 4&339460&197.3&27.5&290.5&33.3&N&900&4795&-59.2&6.0&95&19.3\\
27 &08/14 21 08 13& 9&334520&315.1&6.3&104.9&50.6&Ic&300&3680&80.6&13.8&70&16.3\\
28 &08/14 21 26 22&12&339540&315.1&6.3&106.9&53.4&Ic&300&8489&-24.4&1.0&500&17.1\\
29 &08/14 21 38 23&14&339560&315.1&6.3&109.2&56.1&Ic&300&4534&-20.2&1.0&180&18.1\\
30 &08/15 02 33 54&46&339880&14.0&39.2&33.1&65.8&V&600&4081&-49.9&12.6&125&15.7\\
31 &08/15 03 12 11&50&339920&12.2&40.4&11.5&68.9&Ic&300&9881&-80.8&2.2&1400&15.0\\
32 &08/15 20 17 21& 5&340220&197.3&27.5&290.6&32.7&N&900&4894&61.8&1.2&420&17.0\\
33 &08/15 22 39 15&23&340400&253.0&34.9&300.3&46.6&V&600&5375&-13.0&6.0&350&18.5\\
34 &08/16 00 23 14&32&340490&13.3&39.7&233&45.0&V&600&3441&52.1&14.8&22&18.5\\
35 &08/16 01 29 19&38&340550&17.6&37.4&232&55.0&V&600&7796&-26.5&1.4&1300&16.6\\
36 &08/16 02 59 42&45&340620&17.6&37.4&207&70.0&Ic&300&7538&-19.5&1.4&200&17.7\\
37 &08/16 03 17 49&48&340650&15.0&40.4&184&70.0&Ic&300&7990&5.9&1.2&230&17.6\\
38 &08/16 03 41 51&52&340690&15.0&40.4&184&70.0&Ic&300&7792&3.8&1.4&1900&15.2\\
39 &08/16 03 59 52&55&340720&13.3&39.7&166&69.0&Ic&300&3597&24.2&2.6&30000&11.5\\
40 &08/16 04 05 53&56&340730&13.3&39.7&166&69.0&Ic&300&5679&-32.3&13.0&70&16.3\\
41 &08/16 04 23 56&59&340760&13.3&39.7&166&69.0&Ic&300&7718&-7.5&2.4&240&16.8\\
42 &08/16 04 29 57&60&340770&12.1&38.8&149&67.0&Ic&300&5332&-8.6&2.0&1400&15.1\\
43 &08/16 04 47 56&63&340800&12.1&38.8&149&67.0&Ic&300&4300&40.8&14.4&35&17.0\\
44 &08/15 03 24 13&52&339940&12.2&40.4&5.1&69.3&Ic&300&2371&-55.3&12.0&38&17.1\\
45 &08/13 03 48 27&65&338260&14.0&39.2&0.3&70.5&Ic&300&1820&-39.6&1.4&80&18.6\\
46 &08/13 04 06 29&68&338290&14.0&39.2&349.8&70.2&Ic&300&3274&-16.6&2.0&1080&18.0\\
47 &08/14 01 15 23&40&338930&12.9&39.9&49.0&52.9&V&600&1195&54.1&0.8&20&21.8\\
48 &08/14 03 20 44&58&339110&11.9&39.1&9.1&70.4&Ic&300&2315&-23.0&1.4&170&17.7\\
49 &08/14 04 03 05&65&339180&7.3&37.8&333.8&69.5&Ic&300&3748&13.5&1.0&30&20.1\\
50 &08/14 04 15 07&56&339200&7.3&37.8&328.5&68.1&Ic&300&2539&-48.9&1.4&170&17.8\\
51 &08/15 01 05 45&38&339800&17.3&37.6&54.7&49.4&V&600&3894&-10.9&0.8&70&20.5\\
52 &08/16 04 23 56&59&340760&13.3&39.9&334.4&67.4&Ic&300&2661&44.9&13.0&16&17.9\\
53 &08/16 04 23 56&59&340760&13.3&39.9&334.4&67.4&Ic&300&1399&-29.2&1.0&40&18.9\\
54 &08/13 20 42 50&11&338640&313&6.0&101.8&45.0&Ic&300&2212&60.8&14.6&48&16.6\\
55 &08/13 20 48 51&12&338650&313&6.0&102.6&47.0&Ic&300&2496&37.0&1.6&700&16.1\\
56 &08/15 03 36 15&54&339960&13.3&39.9&6.2&71.0&Ic&300&1420&18.1&2.0&100&18.0\\
57 &08/15 02 33 54&46&339880&14.0&39.2&33.1&65.7&V&600&2983&44.0&13.2&16&19.0\\
\hline
    \end{tabular}
  \caption{Faint tracks recorded during 10-16 August, HST}
\label{tracks}
\end{table*}

 Table~\ref{tracks} summarizes the observational parameters of all of the detected tracks recorded during eight nights of SuprimeCam observation, 8--16 August 2004, HST. Tracks $\#1-\#9$ were reported by I.Iwata. Tracks $\#10-\#43$ were noticed by one of our authors, M.T., while taking a quick look at the observed images onsite.  Tracks $\#44-\#56$ also were later noted by M.T. while reducing the imaged data.   Track $\#57$ was found by M.I. while reducing the meteor images for this paper.  Eventually, this eight-night series of SuprimeCam observations yielded 57 tracks.

A short explanation of each column follows:

Column (1): Sequential number \#

Column (2): Start time of exposure in HST

Column (3): Frame number of the night

Column (4): CCD file number SUPA00nnnnnn (10 files starting from this number constitute the entire SuprimeCam field of view (FOV) covered with 10 CCDs)

Column (5): Right ascension in degrees

Column (6): Declination in degrees

Column (7): Azimuth angle in degrees

Column (8): Elevation angle in degrees

Column (9): Filter band

Column (10): Exposure time in seconds

Column (11): Total track length $l$ in pixels over several CCDs 

Column (12): Position angle $\theta$ of the track

Column (13): Full width half maximum (FWHM) $w$ in arcsec

Column (14): Average peak count per pixel in ADU.

Column (15): Nominal integrated magnitude for the brightest 5 arcmin portion. See section 4 for further details.

\medskip

Figures~\ref{track11} to ~\ref{track39} show typical images among the recorded 57 tracks.  The figure~\ref{track11} labeled \#11\_812\_39\_8, for example, shows about ${1800 \times 1800}$ \rm pixel portion of the image of track \#11 recorded on the 8th CCD of the 39th exposure frame taken on 12 August, HST. To elucidate the tracks, the surface brightness ranges of these images were adjusted individually to give high contrast.

As will be explained later, Figure~\ref{track11} was identified as satellite COSMOS 2363 = ID98077B = NORAD\#25594, flying at a high orbit with a slant range 20,000 km and an orbiting period of 675.73 min. The brightness variation in this frame is visible because of the rotation of the satellite. Similar light modulation along the track was noticed also for $\#5, \#36, and \#45$. 

Figure~\ref{track50} is another satellite DIAMANT\_B-P4 = ID75010A = NORAD\#07654, flying at a low orbit, below 1500 km, with an orbital period 107.69 min.

Figure~\ref{track17} is a Perseid meteor at a distance of about 106 km.  The track is blurred because of the significant defocus effect that occurs when the SuprimeCam  is focused to infinity.

Track \#30 in Figure~\ref{track30} is a sporadic meteor.  This meteor exhibited an outburst in which its luminosity increased by a factor of seven during its passage through the SuprimeCam FOV. Another even fainter track of $\#57$ crossing almost orthgonally to $\#$30 is seen in the lower left corner of this image.

Figure~\ref{track39} is the brightest saturated satellite image among the sample. 
  
 The usual procedures were taken to subtract bias, and flat-fielding the raw frames was performed using the SuprimeCam Data Reduction package SDFRED \citep{yagi02, ouchi04}.

\section{Meteor Identification}
\subsection{Image Width}
An object with a diameter $D$ at a distance $d$ from the telescope, imaged by the SuprimeCam with focal length $f$ focused to the infinity under a typical seeing of $s$ in radian, will have an image size $w$, as given by
\begin{eqnarray}
 w = \sqrt{f^{2} D^{2}d^{-2} + \delta^{2}F^{-2} + f^{2}s^{2}},
\end{eqnarray}

\noindent where the second term denotes the blurred image size due to the defocus $\delta$ of a beam with focal ratio $F=2$ given by 
\begin{eqnarray}
 \delta = f^{2}/(d-f).
\end{eqnarray}

Note that the pixel scale of SuprimeCam is 0.20 arcsec/pixel with 15$\mu$m pixel.  

\begin{figure}
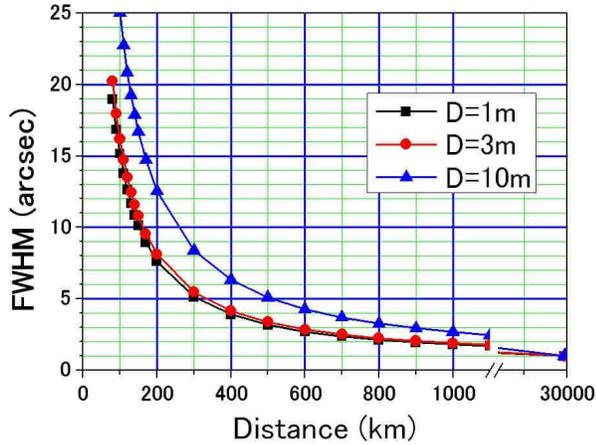

  \begin{center}
    \FigureFile(80mm,70mm){fig6.eps}
  \end{center}
  \caption{FWHM size of the defocused object of diameter $D$ at a distance $d$ from the telescope under a typical seeing size of 0.8 arcsec. Objects smaller than 1m in size are not resolved at 100km distance. \label{sizedist}}
\end{figure}

Figure~\ref{sizedist} shows the calculated FWHM image width (in arcsec) as a function of distance $d$ under typical seeing size of 0.8 arcsec ($s = 4 \times 10^{-6}$ radian) for objects with size D=1m, 3m, and 10m.  At the distance of meteors, 100 km $< d <$ 200 km, the second term of the defocus effect is the leading term; the defocused images of meteors should have an image size larger than 8 arcsec at a distance of less than 200 km.  
Note that size information of an object smaller than 1 m cannot be retrieved from its defocused image width.

\begin{figure}
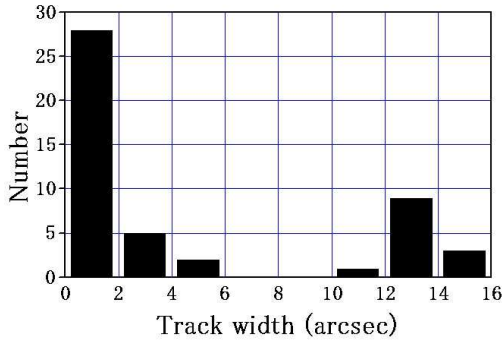

  \begin{center}
    \FigureFile(70mm,60mm){fig7.eps}
  \end{center}
  \caption{Bimodal FWHM size distribution of the detected tracks.\label{sizesnew}}
\end{figure}

Figure~\ref{sizesnew} shows the actual FWHM track width distribution of the 48 measured tracks. The distribution shows a clear separation of the two populations. Thirteen tracks are wider than 10 arcsec and correspond to meteors and the remaining 35 tracks are narrower than 6 arcsec and correspond to satellites and space debris.

\small
\begin{table}
\begin{center}
  \begin{tabular}{rrrrr}
\hline
(1)&(2)&(3)&(4)&(5)\\
Group&Period&Max&RA&Dec\\
\hline
$\gamma$Per&23 Jul.-23 Aug.& 12 Aug.& 47$\degree$& 58$\degree$\\
(N)$\delta$Aqr&7 Jul.-25 Aug.& 12 Aug.&340$\degree$& -5$\degree$\\
(S)$\delta$Aqr&21 Jul.-29 Aug.& 29 Jul.&334$\degree$& -16$\degree$\\
(N)$\iota$Aqr&15 Jul.-20 Sep.&20 Aug.& 328$\degree$& -6$\degree$\\
(S)$\iota$Aqr&15 Jul.-25 Aug.&4 Aug.&334$\degree$& -15$\degree$\\
$\alpha$Cap&15 Jul.-10 Aug.& 31 Jul.& 308$\degree$& -10$\degree$\\
Cas&17 Jul.-15 Aug.&28 Jul.& 15$\degree$& 63$\degree$\\
$\kappa$Cyg&19 Aug.-25 Aug.&  - & 289$\degree$& 55$\degree$\\
\hline
    \end{tabular}
  \caption{Meteor groups active during 10-16 August.}
  \label{groups}
\end{center}
\end{table}

\begin{table*}
\begin{center}
\small
    \begin{tabular}{rllrrrrrrrrlrr}

\hline
(1)&(2)&(3)&(4)&(5)&(6)&(7)&(8)&(9)&(10)&(11)&(12)&(13)\\
ID&ExpStart&EL&$d$&B&$w$&$I_p$&$\theta$&$\delta\theta_{Per}$&$\delta\theta_{Aqr}$&group&$\Psi$&$\Omega$\\
\#&UT h m s&deg&km&&pix&ADU&deg&deg&deg&&deg&$\degree$/s\\
\hline
17 &8/13 14 12 30&69.9&106&Ic&68&38&35.5 &-0.4 &-81.6&Per&29&15.5\\
54 &8/14 06 42 50&45.0&141&Ic&73&48&-60.8&-81.1&29.7&Spo&-&10:\\
19 &8/14 10 42 21&46.5&138&V&61&195&23.5&-13.0&85.2&Spo/P?&-&10:\\
23 &8/14 13 26 47&70.6&106&Ic&67&55&48.0 &11.3&-72.4&Spo/P?&-&10:\\
25 &8/14 14 33 24&70.6&106&Ic&67&120&-66.0&81.2 &2.3&$\delta$Aqr&62&10.2\\
27 &8/15 05 08 13&50.6&129&Ic&69&70&80.6 &60.0 &-6.0&Spo/A?&-&10:\\
30 &8/15 12 33 54&65.8&110&V&63&125&-51.0 &-87.0 &11.5&Spo/A?&-&10:\\
57 &8/15 12 33 54&65.8&110&V&66&16&44.0&8.0&-73.5&Spo/P?&-&10:\\
44 &8/15 13 24 13&69.3&107&Ic&60&38&-55.0 &86.9&4.0&Spo/A?&-&10:\\
34 &8/16 10 23 14&45.0&141&V&74&22&52.1&15.2 &-66.8&Spo/P?&-&10:\\
40 &8/16 14 05 53&69.0&107&Ic&65&70&-32.3 &-69.2 &28.8&Spo&-:&10:\\
52 &8/16 14 23 56&67.0&108&Ic&65&16&45.0&7.9&-74.1&Spo/P?&-&10:\\
43 &8/16 14 47 56&67.0&109&Ic&72&35&41.3 &5.0 &-78.5&Spo/P?&-&10:\\
\hline
    \end{tabular}
  \caption{Meteors recorded during 10--16 August in their order of appearance.}
\label{meteors-1}
\end{center}
\end{table*}

\subsection{Position Angle}
Several meteor groups are known to appear during the time period of the observations as shown in Table~\ref{groups}. Cassiopeids are considered to be a parted family of Perseids.

We performed a cross-check to determine if the position angles of the observed tracks point to any of the radiant points of these meteor groups.  Table~\ref{meteors-1} summarizes the  results of the meteor identification and lists some of the physical parameters that were measured for these 13 meteor tracks.  An explanation of the procedures adopted to derive the physical parameters is given after the short description of each column:

\medskip

Column (1): Sequential ID number

Column (2): Exposure start time in UT(=HST+10h)

Column (3): Elevation angle

Column (4): Distance to the meteor assuming an altitude of 100 km from the ground

Column (5): Filter band used for imaging

Column (6): FWHM $w$ of the track in pixels

Column (7): Average peak count $I_p$ per pixel in ADU

Column (8): Observed position angle $\theta$ of the meteor track

Column (9): Angle deviation $\delta\theta_{Per}$ from $\gamma$Per radiant point

Column (10): Angle deviation $\delta\theta_{Aqr}$ from $(N)\delta$Aqr radiant point

Column (11): Meteor group identification.  Per; Perseids,  $\delta$Aqr; $\delta$Aqr(N), Spo; sporadic. Although possible associations to Perseids and Aquarids are not completely denied, we consider the nine meteors with "Spo/P?", and "Spo/A?" denotation to be sporadic meteors, based on the sporadic meteor event rate as described in the discussion.

Column (12): Angle distance $\Psi$ of the telescope pointing from the radiant point.  

Column (13): Meteor angular speed $\Omega$.  Except for Perseid $\#17$ and Aquarid $\#25$, we adopted a canonical angular speed of 10$^\circ$/s for sporadic meteors to simplify matters.

\medskip

Column (9) of Table~\ref{meteors-1} shows the deviation of the position angle $\delta\theta_{\rm{Per}}$ of the track from the nominal Perseid radiation point. For the daily radiant ephemeris for the Perseid meteor showers, we refer to \citet{arlt03} and \citet{lyyt04}. 

The spread in Perseid radiants was estimated to be about 2--3 degrees wide (cf.\cite{jon79, lind95, mol97}). The analysis of Perseid meteors observed in 1997 (Shigeno et al.; http://www004.upp.so-net.ne.jp/msswg/Ob970812.txt), 
showed that the actual distribution of the radiant points of 23 bright meteors, with magnitudes ranging from 0.5 -- 8.5 mag and mean magnitude 4.76, spread about 3$\degree$, while a fainter group of 14 meteors in the range 6 -- 8.5 mag, with mean magnitude 7.00, spread in a much wider region, 60$\degree$ in right ascension and 20$\degree$ in declination. This suggests that the radiant points of fainter meteors may spread much wider in space.

Track \#17 is aligned to the position angle of the Perseid radiant point within 1$\degree$, and we associate this meteor with  the Perseids. Tracks \#43, 52, 57, 19, 23, and 34 are within 15$\degree$ from the direction of the Perseid radiant point and although some may be Perseids, considering the finding of Shigeno et al.(1997), we prefer to consider them sporadic meteors based on the mean event rates independently evaluated in other fields. 

Column (10) shows the angle deviation $\delta\theta_{\rm{Aqr}}$ to (N)$\delta$Aqr meteor radiation point.  
Here, we identify track \#25, whose track position angle is pointing within $3\degree$ from the Aquarid radiant point, with this group of meteoroids. Again, some of the tracks \#27, 44, and 30, with angle deviation less than $12\degree$, might also be members of this group, or other nearby Aquarid groups, (S)$\delta$Aqr, (N)$\iota$Aqr, or (S)$\iota$Aqr, as shown in Table~\ref{groups}.  Note that the position angle matched best to (N)$\delta$Aqr for all four tracks.  We interpret that all of the meteors except track \#25 are largely sporadic, according to the same reasoning.

The remaining two, \#40 and 54, did not match the radiant points of the groups listed in Table~\ref{groups} and are considered here to be secure sporadic meteors. 
The number of sporadic meteors in our sample is therefore at most 11.

Figures~\ref{mete1-7} and ~\ref{mete8-13} show the identified meteor tracks recorded in several contiguous CCDs.  Contrast levels were adjusted individually to make the tracks visible. Some of the meteors showed significant changes in their luminosity during their passage of the SuprimeCam FOV. The luminosity profile of these images is discussed in section 4.2.  Note that doubly split images of meteors $\#27$ and $\#43$ are due to the defocused central obscuration of the telescope.

The distance $d$ to the meteor is given in column (4) of Table~\ref{meteors-1} using the relation $d=H$/sin(El), where the luminous altitude $H$=100 km was assumed. The angular speed, $\Omega$, of the meteor entering the luminous height at a velocity $v$ with an angular separation $\Psi$, from the radiant point is given in column (13) by
\begin{eqnarray}
  \Omega = v\  \rm{sin}(\Psi) / d ,  
\end{eqnarray}
\noindent for a Perseid and an Aquarid meteor, respectively.  For other sporadic meteors, we adopted a common canonical value of 10 degrees/sec, for simplicity.

\medskip

\begin{figure*}
  \begin{center}
    \FigureFile(130mm,100mm){fig8.eps}
  \end{center}
  \caption{Meteor tracks\label{mete1-7}}
\end{figure*}

\begin{figure*}
  \begin{center}
    \FigureFile(130mm,100mm){fig9.eps}
  \end{center}
  \caption{Meteor tracks \label{mete8-13}}
\end{figure*}

\begin{table*}
\begin{center}
\small
    \begin{tabular}{rllrrrrrrrrlrrrr}
\hline
(1)&(2)&(3)&(4)&(5)&(6)&(7)&(8)&(9)
\\
ID&ExpStart&B&$w I_p$&group&$m_{5'}^{nom}$&$\Omega$&$m_{vr}$&$D_{col}$\\
\#&UT h m s&&ADU&&mag&$\degree$/s&mag&mm\\
\hline
17 &8/13 14 12 30&Ic&2584&Per&16.9&15.5&4.9&-\\
54 &8/14 06 42 50&Ic&3504&Spo&16.6&10:&5.0:&-\\
19 &8/14 10 42 21&V&11895&Spo/P?&16.4&10:&4.2&10.3\\
23 &8/14 13 26 47&Ic&3685&Spo/P?&16.5&10:&5.0&-\\
25 &8/14 14 33 24&Ic&8040&$\delta$Aqr&15.7&10.2&4.1&-\\
27 &8/15 05 08 13&Ic&4830&Spo/A?&16.3&10:&4.7&-\\
30 &8/15 12 33 54&V&7875&Spo/A?&15.7&10:&4.6&6.7\\
57 &8/15 12 33 54&V&1056&Spo/P?&19.0&10:&6.8&2.4\\
44 &8/15 13 24 13&Ic&2280&Spo/A?&17.1&10:&5.5&-\\
34 &8/16 10 23 14&V&1628&Spo/P?&18.5&10:&6.4&3.9\\
40 &8/16 14 05 53&Ic&4550&Spo&16.3&10:&4.8:&-\\
52 &8/16 14 23 56&Ic&1040&Spo/P?&17.9&10:&6.4&-\\
43 &8/16 14 47 56&Ic&2520&Spo/P?&17.0&10:&5.4&-\\
\hline
    \end{tabular}
  \caption{Physical characteristics of meteors recorded.}
  \label{photometry}
  \end{center}
\end{table*}

\section{Photometry}

\subsection{Absolute photometry and nominal CCD magnitude}

We performed absolute photometry of the CCD frames by imaging twice in each night two of the following four photometric standard star fields, SA110, SA92, PG2332+055, and MARK\citep{land92}.   The photometric zero points of V-band and Ic-band were established by 14 to 24 standard stars in the magnitude range $11.8 \le V \le 16.3$ and $10.7 \le Ic \le 15.3$ available in each shot. 

For example, the Ic-band zero point of the CCD frame SUPA00338307 for $\#17$ and SUPA00339230 for $\#25$ were 33.39 and 33.39 mag, respectively.  The V-band zero point of SUPA00339888 for $\#30$ and SUPA00338900 for $\#19$ were 34.47 and 34.48 mag respectively.  The deviation of these zero points was well less than 0.1 mag.

Table~\ref{photometry} summarizes the photometric properties of the 13 meteoroids.   Definition of columns (1) to (5) are common to those in Table~\ref{meteors-1}. Column (6) denotes the nominal integrated magnitude, $m_{5'}^{nom}$, of the brightest 5 arcmin portion of the track, as derived by comparing the integrated ADU counts, 
\begin{eqnarray}
F_{5'}^{nom} = w I_p \times 1500, 
\end{eqnarray}

\noindent
where the factor 1,500 is the number of pixels along the 5 arcmin portion of the track with those of standard stars for longer integration time.   
For example, the Perseid meteor \#17 has $wI_p$ = 2584ADU/pixel, and the total ADC $F_{5'}$ within the brightest 5 arcmin track length is ${3.9 \times 10^6}$.  

The nominal magnitudes are given by
\begin{eqnarray}
V_{5'}^{nom} = -2.5 log(F_{5'}^{nom})+34.47, 
\end{eqnarray}

and 
\begin{eqnarray}
I_{5'}^{nom} = -2.5 log (F_{5'}^{nom})+33.39.
\end{eqnarray}

Note that this is a \it{nominal} \rm magnitude because CCD photometry for temporary luminous objects like meteors gives apparently fainter nominal magnitudes for CCD images with longer exposure times. 

To characterize the meteor magnitude more quantitatively, taking into account its luminous time, we introduce a possible definition of meteor magnitudes in column (8) by calculating/assuming the meteor angular velocity. This is discussed in sections 4.2 and 4.3.   The physical meaning of column (9) is discussed in section 5.1.

\medskip

Column (1): Sequential ID number

Column (2): Exposure start time in UT(=HST+10h)

Column (3): Filter band used for imaging

Column (4): $w I_p$ in ADU per unit pixel along the track

Column (5): Meteor group identification

Column (6): Nominal magnitude, see text above

Column (7): Meteor angular speed $\Omega$.
For sporadic meteors, we assumed a canonical angular speed of 10$^\circ$/s to give the estimated values for columns (8) and (9).

Column (8): Video-rate magnitude $m_{vr}$ defined as $V_{vr}$ or $I_{vr}$.  This magnitude corresponds to the spatial integration of all of the photons from the portion of the track length traversed within 0.033 sec.  See section 4.2. Note that this includes, for the V-band, the afterglow contribution from [OI]5577 lasting about 1 sec. 


Column(9): Calculated maximum diameter of collisionally excited meteors column $D_{\rm{col}}$.  See section 5.1 for further details.

\subsection{Video-rate magnitude}
The number of 0.2 arcsec pixels that span a video frame of 0.033 sec, for a meteor with angular speed $\Omega$ deg/s is $3600 \times 0.033 \Omega / 0.2 = 590 \Omega $ pixels. The total ADC for the brightest observed part of the track $F_{vr}$ in 33 msec is then
\begin{eqnarray}
F_{vr} &=&  w I_p \times 590 \times \Omega \nonumber \\
  &=&  F_{5'}^{nom} \times \Omega \times 590 / 1,500,
\end{eqnarray}

\noindent                          
where $wI_p$ is the average ADU count integrated perpendicularly across the track over the blurred image of the meteor given in column (4) of Table~\ref{photometry}.  For example, the Perseid meteor \#17 has $wI_p$ = 2584ADU/pixel and an angular speed of $\Omega = 15.5\degree$/sec at an elevation angle $EL = 70\degree$. 
The total ADC $F_{33}$ within a video-rate frame, corresponding to a track length of 0.52$\degree$, or 9,200 pixels, would then be about ${2.4 \times 10^7}$.  



We define the video-rate magnitudes $V_{vr}$ and $I_{vr}$ by
\begin{eqnarray}
V_{vr} = V_{5'}-2.5 log(\Omega)-9.64,
\end{eqnarray}

and 
\begin{eqnarray}
I_{vr} = I_{5'}-2.5 log (\Omega)-8.88,
\end{eqnarray}

\noindent
where the offsets of 9.64 and 8.88 correspond to the magnitude differences 2.5 log(600/0.033$\times$590/1500) and 2.5 log(300/0.033$\times$590/1500), respectively.

These values are shown in column (8) of Table~\ref{photometry}.  Note that the actual video-rate magnitude could be somewhat fainter than the values given in column (8) as this magnitude is calculated assuming that the brightness remains at a constant maximum value throughout the 0.033sec.



\subsection{Magnitude Linking to Visual Estimation}
Note, however, that the magnitude $m_{vr}$ given here cannot be linked well to the magnitudes reported by most of eye-witness observers.  
Quantitative comparison between the "integrated" meteor brightness and background stars is not straightforward because the effective length of integration time corresponding to the eye response time and the residual memory time has not been well calibrated.  
The eye-witness magnitude may give a much fainter value than the currently introduced photometric definitions of meteor magnitude, $m_{vr}$. Another physical factor to be considered in calibrating magnitudes is the time profile of meteor radiation.  Because a small, but significant, fraction of the photons in the V-band is emitted at the [OI] 5577A forbidden line that lasts 1--3 sec, the actual duration of emission recorded in the present observation would be longer than the adopted video-rate and hence the actual magnitude corrected for this afterglow duration time, $m_{cor}$, could be at most 2--3 mag fainter than $V_{vr}$.   In the $I_c$ band, where the OI 7774A and $N_2$ bands are dominant and no significant forbidden line contributes, such an afterglow effect would be negligible.

Therefore, most of the currently observed meteors could possibly be as faint as 7 --8 mag to visual observers.  However, we will not pursue a rigorous quantitative calibration between eye estimate magnitudes and the photon counting magnitude in this paper.  

\subsection{Luminosity Profiles}

\begin{figure*}
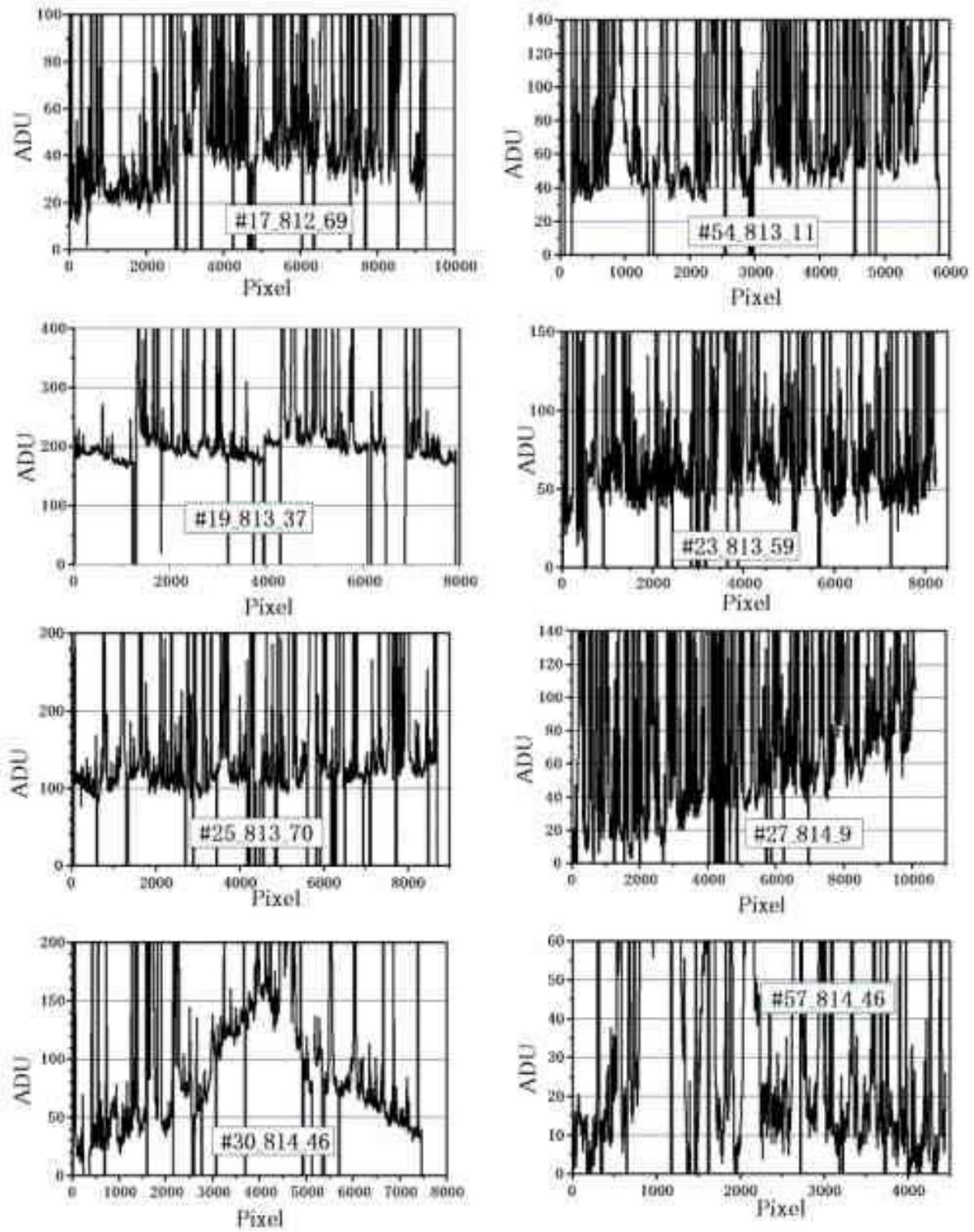

  \begin{center}
    \FigureFile(160mm,240mm){fig10.eps}
  \end{center}
  \caption{Light profiles of meteors \label{prof1-8}}
\end{figure*}

\begin{figure*}
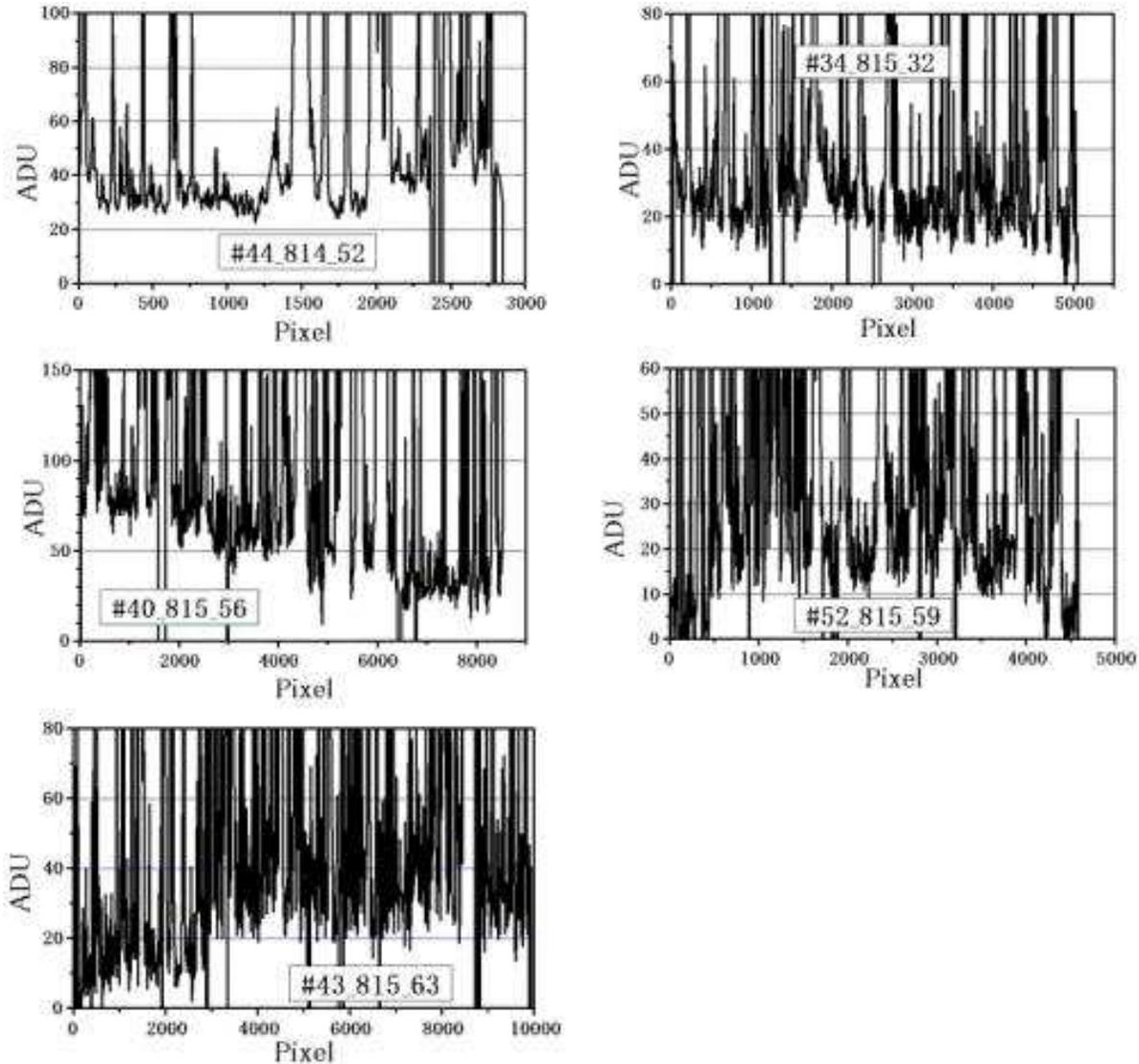

  \begin{center}
    \FigureFile(160mm,180mm){fig11.eps}
  \end{center}
  \caption{Light profiles of meteors (continued) \label{prof9-13}}
\end{figure*}

Figures~\ref{prof1-8} and ~\ref{prof9-13} present a luminosity profile scanned along the track of all of the recorded meteors.  The ADU count averaged over the FWHM is shown in these plots. Bright spikes in the profile are due to stars and/or hot pixels and the lower envelope of the profile indicates the meteor luminosity.  Some of the meteors, for example, $\#19$ and $\#23$, show fairly constant brightness.  However, the meteor $\#30$ flared up by a factor of 7 in brightness and meteor $\#27$ showed a gradual change in luminosity by a factor of 3 within the recorded frame.  
The FWHM track length for the flare-up of meteor $\#30$ is about 500 arcsec (= 2500 pixels), corresponding to an elapsed time of 14 msec.  This event could be employed to constrain the ablation process through the use of an appropriate model.

\section{Discussion}
\subsection{Rate of Sporadic Meteors}
To evaluate the rate of sporadic meteors, we examined 209 SuprimeCam frames (2090 CCD images) obtained for the Subaru Deep Field (SDF) during 2001--2004, which are publicly available in the Subaru data archive.  173 frames are in the $I'$ band with an exposure time of 210 to 300 sec, and 46 frames are in the $V$ band with an exposure time 720 to 900 sec.  Most of the exposures were taken in March, April, and May.  The total exposure time of these frames amounted to 21.6 hours, somewhat longer than the  19 hours for the current M31 observation.  A total of 29 meteor events was confirmed in these SDF images. The SDF is in the constellation of Coma, where a modest meteor group was expected around 19 January. However, considering that the SDF observations were made much later than this shower period and that no other prominent meteor showers occurred during this time, we regard most of the SDF meteors as sporadic. Since we have no information that allows us to subtract possible shower events in SDF meteors, we retained all of the 29 SDF meteors as well as 13 M31 meteors in the current observation to compare their general event rates. The average event rate of SDF meteors thus obtained is $1.96\pm0.63$ times larger than, but comparable to what we obtained in the M31 field, supporting the conclusion that most of the meteors observed are also sporadic. We could not determine whether the factor of two difference in the event rate is significant.

\subsection{Cross Section of the Collision Column}

As shown in Figure~\ref{sizedist}, the current defocused observation cannot resolve a source size smaller than about 1m.  However the following consideration gives a crude size estimate of the column cross section where the meteor collided with atmospheric atoms.

Adopting the system's photon detection efficiency of $50\%$ by taking the optics throughput, filter transmission, and CCD quantum efficiency into account, and applying the ADC conversion factor of 2.6 e/ADC for the SuprimeCam, we can evaluate the integrated photon counts.

For example, the meteor track $\#19$ produced a total of $5.7 \times 10^{8}$ photons received in the V-band by the Subaru Telescope during its 33 msec travel of $L=2$ km length at 60km/sec and at about 100km in height, corresponding to a distance of 138km from the telescope. 
Assuming isotropic emission, one can derive the total number of photons emitted by this meteor during the 33msec as $N_{photon} = 2.6 \times 10^{18}$ photons.  The average density of neutral oxygen atoms at a 100km height is about $n_{OI} = 10^{18}$ m$^{-3}$.   As a fraction $\eta$ of all of the photons are emitted in [OI] 5577 by collisional excitation, this requires the effective cross section $S$, given by the following equation; 
\begin{eqnarray}
\eta N_{photon} = n_{OI} S L.
\end{eqnarray}

A typical meteor spectrum gives a rough estimate of the fraction of [OI] contribution in the V-band at about $\eta=0.1$.  By inserting the observed numbers for meteor $\#19$, we obtain $S=8.3\times10^{-5}$m$^2$, corresponding to an effective diameter of the collision column of $D_{col}=10.3$mm. The diameter of the collision column thus calculated is given for four meteors observed in the V-band in column (9) of Table~\ref{photometry}, which is in the range of 2.4mm $\le D_{col} \le 10.3$mm .

Note that this value represents the diameter of the column where the main body of the meteor collided with atmospheric atoms, molecules, and electrons, and where the successive cascade collisions of these particles with the neutral oxygen atoms took place. These collisions excited the neutral oxygen atoms to the energy level, which released the subsequent forbidden line [OI] 5577.

On average, the actual emission of the forbidden line occurred 0.7 sec after the collisional excitation. Therefore, the excited oxygen atoms were dispersed by thermal motions up to few hundred meters away from the original collision column before producing the forbidden-line radiation.  The size of the radiating zone of the forbidden line, therefore, would be as large as few 100 m.

Note that this should be the size observed in [OI]5577 monochromatic images of meteors. The width of meteor images recorded in normal broadband frames represents the size of the hot column, which is much narrower than the [OI] wake column and would be about 1 cm -- 1 m.   This is consistent with the independent result of two station, short baseline observation of 34 faint meteors using intensified CCD cameras \citep{kai04}.

\subsection{Radiation Efficiency and Meteor Mass}

The kinetic energy $E_k$ of a meteor particle of mass $m$ with relative velocity to the Earth $v$ is given by 
\begin{eqnarray}
E_k = m v^{2}/2 . 
\end{eqnarray}

Again, by assuming isotropic radiation, a photon detection efficiency of $50\%$, and an ADC conversion factor of 2.6 e/ADC, we can estimate the total number of photons $N_{ph}$ emitted by the meteor during, for example, the 33msec from the video-rate ADC count $F_{vr}$ measured by the Subaru Telescope. 
\begin{eqnarray}
N_{ph}=4\pi d^{2} F_{vr} / \pi R^{2} \times 2.6 \times 2,
\end{eqnarray}

\noindent 
where $d$ is the distance to the meteor and $R$=4.1m is the radius of the Subaru primary mirror.

The total photon energy emitted is then
\begin{eqnarray}
E_p = h\nu N_{ph},
\end{eqnarray}

\noindent where $h\nu$ is the mean energy of a photon, and the conversion efficiency $\epsilon$ of thermalized energy into photons is therefore, 
\begin{eqnarray}
\epsilon = E_p / E_k.
\end{eqnarray}

Or, conversely, if we assume $\epsilon=0.002$ \citet{camp04}, we can estimate the mass of the meteor from its magnitude as
\begin{eqnarray}
m = 2 E_p / v^{2}\epsilon
\end{eqnarray}

For meteor \#19, for example, $E_p$= 0.94J, for an average $V$-band photon of $3.6\times 10^{-19}$J.   The corresponding mass would therefore be $m= 1.92/(30,000)^2/0.002=1.1\times 
10^{-6}$kg= 1.1 mg

\subsection{Faint meteor population and [OI] imaging}

 \citet{pawl01} observed very faint Leonid meteors in 1999 using an intensified CCD system mounted on a 3m liquid mirror telescope (LMT). Although their observation was sensitive down to 18 mag, the integrated magnitude along the track ranged from 5 to 10 mag, corresponding to a meteor mass of 100 to 1 $\mu$g. They estimated that the number flux of these small Leonid meteors was about 1/hour/km$^2$ perpendicular to the Leonid stream.
Pawlowski's LMT observation with IICCD, covering a 0.28$\degree$ field, enabled detection of 140 non-Leonid events per hour, which is astonishing.  Our present observation was not as deep as their detection sensitivity due to the defocus effect of SuprimeCam.  Dedicated meteor imaging with SuprimeCam focused to about 150 km distance would yield about a 3 mag gain in sensitivity and would provide important information on the faint population of meteors.

 SuprimeCam Perseid radiant point imaging using an intermediate band filter IB550 to make [OI] 5577 line imaging would be useful to study the collisional excitation profile of the meteor events and their successive line emitting time profile as shown by the drifting meteor wakes.  Two-dimensional images of each wake will show profiles of collisional excitation and their drifting emission.  SuprimeCam should be offset by about 1.5mm to focus at 150 km distance, which is close to, but still within, the limit of the top-unit travel dynamic range.  Such an observation will also give clues to the faint population of meteoroids and would be worth planning.

\begin{figure*}
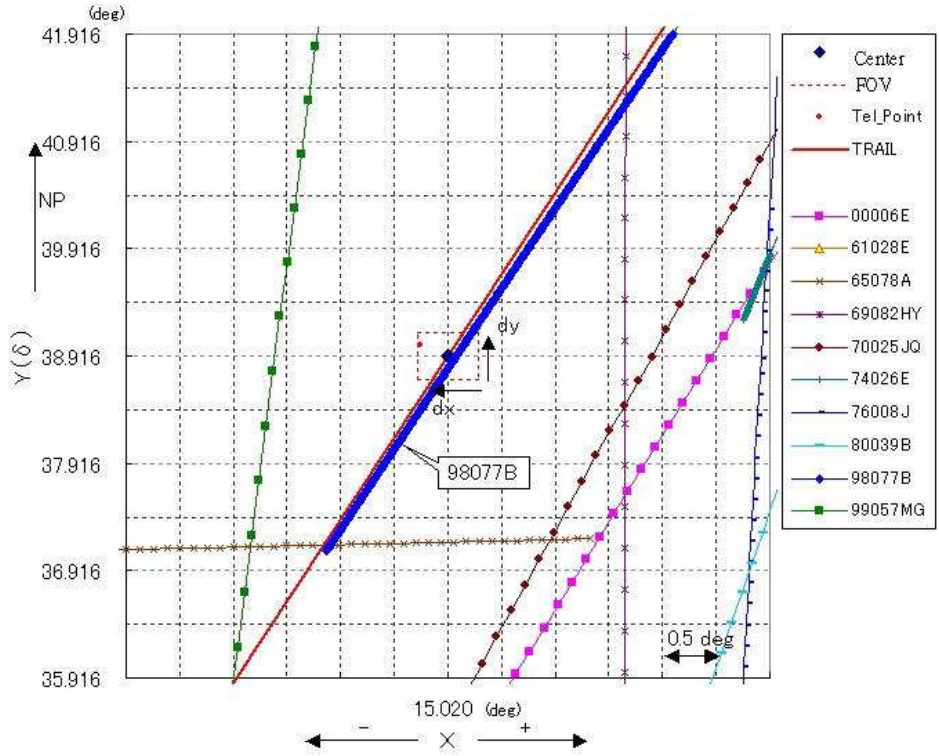

  \begin{center}
    \FigureFile(130mm,100mm){fig12.eps}
  \end{center}
  \caption{Candidate orbiting satellites and debris for Track \#11}\label{orbit11}
\end{figure*}

\begin{figure*}
  \begin{center}
    \FigureFile(130mm,100mm){fig13.eps}
  \end{center}
  \caption{Candidate orbiting satellites and debris for Track \#12}\label{orbit12}
\end{figure*}

\begin{table*}
\begin{center}
\small
\begin{tabular}{llllllrlrrrr}
\hline
(1) & (2) & (3) & (4) & (5)  \\
&(6)& (7) & (8) & (9) & (10) & (11)\\
\hline
\#&Obs Start UTC&ID&NORAD&Object Name\\
&Semi-major axis&Eccentricity&Apogee&Perigee&Period& SlantRange\\
\hline
&UTC&&&\\
&(km)&&(km)&(km)&(min)&(km)\\
\hline\hline
1&08/11 14:03:58&99057F&\#26119&CZ-4\_DEB\\
&7487.2&0.051&1493.6 &724.6&107.53&1000\\ 
2&08/11 14:14:59&75052DF&\#21385&DELTA\_1\_DEB\\
&7315.6&0.018 &1069.7&805.3&103.86&1000 \\
3&08/12 11:53:08&62029B&\#00341&DELTA\_1\_R/B\\
&9663.4&0.242& 5621.7&948.8&157.52&6000\\ 
4&08/12 12:04:08&82110D&\#13658&SBS\_3\_R/B\\
&23624.4&0.716& 34169.0&323.5&602.08&25000\\ 
6&08/12 13:10:13&83059E&\#14136&PALAPA\_B1\_R/B\\
&23080.4& 0.712&33139.3&265.3&581.41&14000\\ 
9&08/12 13:32:14&63014M &\#02362&WESTFORD\_NEEDLES\\
&9999.5 &0.019&3808.0&3434.8&165.92&4000 \\ 
7&08/12 13:32:14&72097B&\#06306&DELTA\_1\_R/B\\
&7683.2&0.027& 1512.4&1097.8&111.78&1200\\
11&08/13 10:34:14&98077B&\#25594&COSMOS 2363\\
&25507.5&0.001& 19160.6&19098.1&675.73&20000\\
15&08/13 13:54:27&97015D&\#26764&SL-6\_R/B(2)\\
&26332.4&0.630& 36532.7&3375.7&708.80&13000\\
16&08/13 14:06:29&93002A&\#22309&MOLNIYA\_1-85\\
&25997.2&0.739 &38825.3&412.9&695.31&12000\\
46&08/13 14:06:29&70025AM&\#04653&THORAD\_AGENA\_D\_DEB\\
&72727.4&0.003&873.1&825.4&101.93&-\\
22&08/14 11:37:29&89001C&\#19751&COSMOS 1989\\
&25503.2&0.001& 19160.4&19089.7&675.56&20000\\
24&08/14 13:44:55&63047F&\#00700&ATLAS CENTAUR 2 DEB\\
& 7502.4&0.073&1671.7&576.7&107.69&1500\\
50&08/14 14:15:07&75010C&\#07654&DIAMANT\_B-P4\\
& 7298.4&0.018&1049.7&790.9&103.42&-\\
51&08/15 11:05:00&85118R&\#23294&SL-12\_DEB\\
& 7502.4&0.073&1671.7&576.7&107.69&-\\
37&08/16 13:17:49&94011E&\#23003&COSMOS\_2272\\
&7780.3&0.001&1411.1&1393.2& 113.84&- \\
53&08/16 14:23:00&77121X&\#10580&COSMOS\_970\_D\\
& 7478.1&0.007&1151.4&1048.5&107.27&- \\
\hline
\end{tabular}
\end{center}
\caption{Information of identified artificial space objects based on Two Line Elements released by NASA/GSFC Orbital Information Group (Data on 08/11/2004 -- 08/16/2004).}
\label{artificial}
\end{table*}

\section{Satellites and Space Debris}
 Some of the tracks were promptly identified by K.Y. as artificial satellites by cross-checking with the NORAD Two-Line-Element database of registered orbit data for 8,888 satellites as of 15 August 2004, which was provided by Mike McCants via Yokohama Kagaku-kan.  The possible satellites that pass the $1\degree\times 1\degree$ FOV centered at the telescope pointing during the 5 minute before the start of exposure and 5 minute after the end of exposure were listed up and compared to the recorded track for its position angle.  Satellite tracks candidates thus identified are given their NORAD number in the fourth column of Table~\ref{artificial}.  No check was performed to determine whether the satellite was in a sunny region and thus reflected sunshine.   

A separate, more extensive check was performed by C. H. at JAXA based on the Two-Line-Element data released by the NASA/GSFC Orbital Information Group.  For example, Figure~\ref{orbit11} shows the tracks of artificial objects that came close to the observed field of track$\#11$ at the time of exposure.  In this case, the artificial object, ID98077B, was confirmed to pass the SuprimeCam FOV during the 600 sec exposure time to take the image $\#11\_812\_38$ along the direction of the observed track.  However, the track found in Figure~\ref{orbit12}, for example, did not match to any of the 8,888 listed artificial bodies. 

As shown in Table~\ref{artificial}, seventeen tracks among the 44 non-meteor tracks coincided with recorded artificial satellites or space debris.  Twenty-seven tracks, however, were not identified with an object in any available catalog.  Some of these might be uncatalogued space debris objects. A possible error in orbital elements due to the secular changes of orbit, especially for low orbits objects, can also lead to cross-identification failure.  

\section{Conclusion}
The Subaru prime focus camera, when focused to infinity, gives a conspicuously defocused image of meteors. The methods to discriminate meteors from artificial satellites/space debris were discussed.  Serendipitous detection of 13 meteors during the later phase of the Perseid shower period was reported, but only 1 was securely identified as a member of the Perseids.  Another meteor was ascribed to the Aquarids, but most of the remaining meteors were likely sporadic ones. Photometric magnitudes of these meteors were measured from the CCD frames by introducing a definition of video-rate magnitude to integrate flux along the tracks.  Problems in linking eye-estimated magnitudes by visual observers to these photometric CCD magnitudes were discussed.

Simple calculation of the number of collisional excitations required to account for the [OI] 5577 forbidden line flux provides a new method to evaluate the diameter of the collision columns of these meteors. In this way, we derived, for the first time, that the sizes of the collision columns of these meteors are approximately a few mm in diameter. Considering the relatively long lifetime (0.7 sec) for releasing the [OI] 5577 emission, the actual column size of the [OI] 5577 line-emitting region would be few 100m in diameter due to the thermal diffusion of excited oxygen atoms before releasing the forbidden line photons.

\section{Acknowledgement}
All of the meteors found were in images taken for the M31 study made by M. Chiba of Tohoku University, P. Guhathakurta of the University of California, and two of the current authors, M. I. and M. T. We are grateful to M. Chiba and P. Guhathakurta for allowing us to investigate meteor samples in the observed data. I.Iwata of NAOJ promptly provided information of tracks \#1--\#9 in Table~\ref{artificial} found during an observation preceding ours.  Discussions with N.Takato of NAOJ on magnitude definition and T.Hashimoto of University of Tokyo on line emission mechanisms were stimulating and useful.  M.Horii of JAXA assisted in the orbital identification of artificial space objects.


\end{document}